\documentclass[sigplan,screen]{acmart}
\AtBeginDocument{%
	\providecommand\BibTeX{{%
			\normalfont B\kern-0.5em{\scshape i\kern-0.25em b}\kern-0.8em\TeX}}}

\setcopyright{acmcopyright}
\copyrightyear{2023}
\acmYear{2023}
\acmDOI{XXXXXXX.XXXXXXX}

\acmConference[DAC'24]{Proceedings of the 61th ACM/IEEE Design Automation Conference (DAC)}{June 23--27,
	2024}{San Francisco, CA, USA}
\acmPrice{15.00}
\acmISBN{978-1-4503-XXXX-X/18/06}

\usepackage{multirow}
\usepackage{setspace}
\begin{document}
	
	\title{iPREFER: An Intelligent Parameter Extractor based on Features for BSIM-CMG Models}
	
	 %%used to denote shared contribution to the research.
	\author{{Zhiliang Peng\textsuperscript{1}, Yichen Wang\textsuperscript{1}, Zhengwu Yuan\textsuperscript{1}, Xiangshui Miao\textsuperscript{1}, Xingming Liu\textsuperscript{2}, Yong Dai\textsuperscript{2}, and Xingsheng Wang* \textsuperscript{1}}\\
		\textsuperscript{1}\small{School of Integrated Circuits, HUST, Wuhan-430074, China }
		\textsuperscript{2}\small{SMiT Group Fuxin Technology Limited, Shenzhen-518054, China }}

%%	\author{Anonymous authors}	
%%	\affiliation{
%%		\institution{Paper under double-blind review} \country{country}
%%	}
%%	\renewcommand{\shortauthors}{Anonymous Author, et al.}

	%%\renewcommand{\shortauthors}{Zhiliang Peng and Xingsheng Wang, et al.}
	
	\begin{abstract}
		This paper introduces an innovative parameter extraction method for BSIM-CMG compact models, seamlessly integrating curve feature extraction and machine learning techniques. This method offers a promising solution for bridging the division between TCAD and compact model, significantly contributing to the Design Technology Co-Optimization (DTCO) process. The key innovation lies in the development of an automated IV and CV curve feature extractor, which not only streamlines the analysis of device IV and CV curves but also enhances the consistency and efficiency of data processing. Validation on 5-nm nanosheet devices underscores the extractor’s remarkable precision, with impressively low fitting errors of 0.42\% for CV curves and 1.28\% for IV curves. Furthermore, its adaptability to parameter variations, including those in Equivalent Oxide Thickness (EOT) and Gate Length (LG), solidifies its potential to revolutionize the TCAD-to-compact model transition. This universal BSIM-CMG model parameter extractor promises to improve the DTCO process, offering efficient process optimization and accurate simulations for semiconductor device performance prediction. It represents a significant step forward in supporting the innovation and progress in semiconductor design and process optimization.
	\end{abstract}
	



	%%\received{20 February 2007}
	%%\received[revised]{12 March 2009}
	%%\received[accepted]{5 June 2009}
	
	\maketitle
	
	\section{Introduction}
	With the development of Moore's Law\cite{moore}, the feature dimensions of transistors continue to decrease, leading to increasingly intricate processes and structures\cite{1ntelfinfet,finfet,2finfet,3finfet}. Factors such as random variability\cite{variabl1,DTCO}, quantum effects\cite{liangzi}, and parasitic effects impose significant limitations on the design space of circuits\cite{jishen}. In response, the Design-Technology Co-Optimization (DTCO) methodology has been proposed\cite{DTCO1}. As illustrated in Figure 1, this figure depicts the collaborative optimization approach of critical components such as compact model in the realm of DTCO\cite{dtcoFLOW}.
	\vspace{-10pt}
	\begin{figure}[h]
		\centering
		\includegraphics[width=0.8\linewidth]{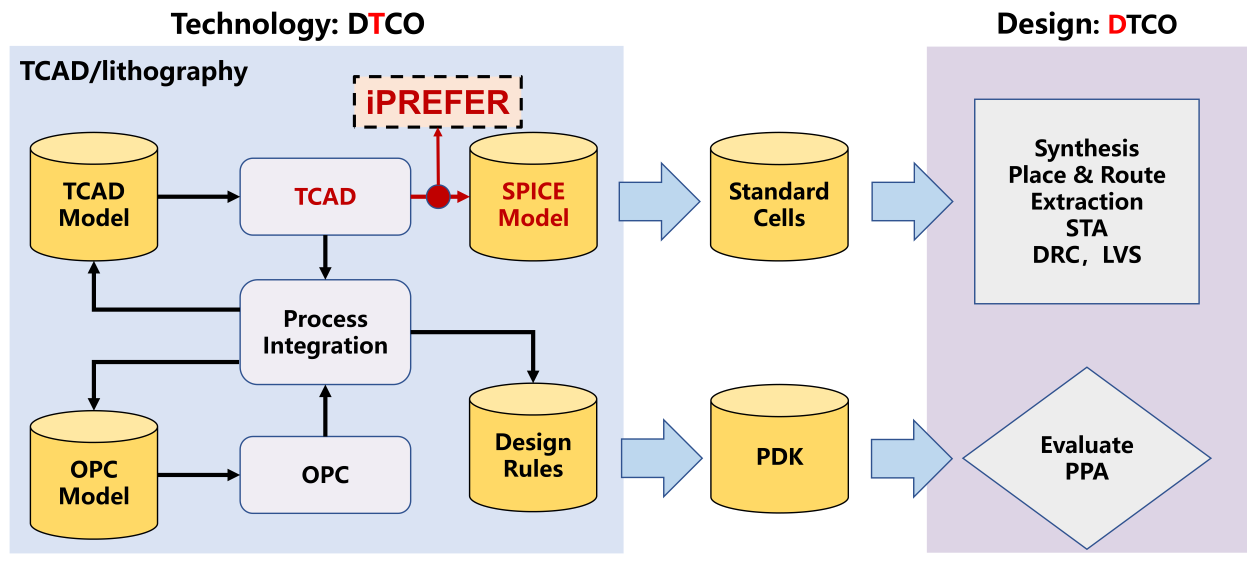}
		\caption[]{\textbf{\small{Simplified DTCO Workflow}}}
		\label{fig:dtco}
	\end{figure}
	\vspace{-10pt}

	\begin{figure*}[h]
		\setstretch{0.5} 
		\centering
		\includegraphics[width=0.85\linewidth]{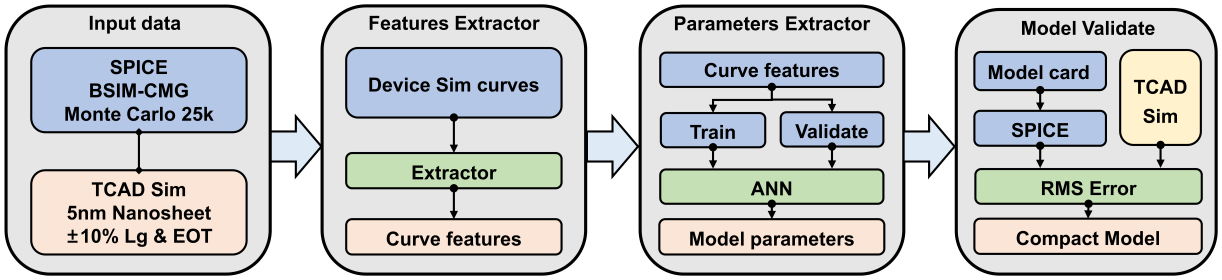}
		\caption{\textbf{\small{Main Workflow Diagram of iPREFER}}}
		\Description{the 3D structural diagram and cross-sectional view of a 5nm nanosheet.}
	\end{figure*}
 
	Compact models act as a crucial link between device fabrication and circuit design. Traditional methods for extracting compact device models are time-consuming, requiring days of expert input for a single model. This delays the rapid transmission of process splits to the design phase.
	While attempts using Artificial Neural Networks (ANNs) aim to expedite the transition from devices to designs, models lacking a physical representation are far from replacing compact models\cite{annmodel1,annmodel2}. Other methods utilizing genetic algorithms for automated parameter extraction face challenges of high computational complexity and time consumption, especially with numerous parameters\cite{tican1}. Researchers have explored integrating ANNs, deep learning models, and optimization techniques for autonomous adjustment of compact model parameters, enhancing extraction efficiency. However, these methods often require specific measurement voltage bias points, limiting input data flexibility in terms of size and patterns\cite{ticandl}. Some studies show the potential of deep learning models to predict parameter values without relying on specific bias points\cite{dlpe}. However, these models usually require input data with fixed-size and resolution , lacking flexibility for optimal performance with unmatched data. Although a proposed cascaded structure aims to directly predict compact model parameters from CV and IV curves of any size\cite{dlfpe}, it faces challenges due to its complex network structure, substantial data requirements, and lower parameter extraction accuracy. 

	To address the aforementioned challenges, this paper proposes a universal, flexible, and precise compact model parameter extraction method based on feature extraction and machine learning. The approach aims to extract a comprehensive set of features from Cgg-Vg and Ids-Vgs curves, utilizing these curve features to accurately and reliably predict BSIM-CMG parameters without the need for iterative processes. Our primary contributions are as follows:
	
	\begin{itemize}
		\item We introduce an intelligent parameter extraction tool with exceptional precision, surpassing other algorithms in terms of network scale, required dataset size, training speed, and parameter extraction accuracy.
		\item We pioneer the application of AI-based compact model parameter extractors within the overall Design Technology Co-Optimization (DTCO) process.
		\item Leveraging a 5nm nanosheet device fabrication platform built on TCAD tools\cite{tcad}, we validate the developed parameter extractor's exceptional precision through device TCAD simulation curves.
	\end{itemize}

	\section{PRELIMINARIES}
	This chapter will commence by introducing the 5nm nanosheet TCAD device fabrication platform, followed by an overview of the overall architecture of the intelligent parameter extraction tool.

	\subsection{5nm Nanosheet TCAD Device}
	
	This paper utilizes a horizontally stacked nanosheet field-effect transistor at the 5nm technology node as the test platform, extracting the corresponding compact model based on the BSIM-CMG  model\cite{bsim}. The structural design of the nanosheet transistor device is derived from experimental data published by IBM in VLSI 2017\cite{nanosheet}. Figure 3 presents the 3D structure and cross-sectional view of the device. The device consists of three layers of horizontally stacked nanosheets, with a gate length of 12 nm, and each nanosheet layer has a thickness of 5 nm. Table 1 provides detailed structural and process parameters for the device.	To achieve precise compact model extraction for the nanosheet transistor device, the target curves used during parameter extraction are sourced from calibrated TCAD data.
	\begin{figure}[h]
		\setstretch{1} 
		\centering
		\includegraphics[width=0.55\linewidth]{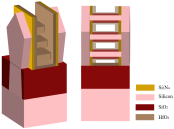}
		\caption{\textbf{\small{5nm Nanosheet's 3D Structural Diagram and Cross-sectional view.}}}
		\vspace{-15pt} 
		\Description{ 3D Structural Diagram and Cross-sectional View of the 5nm Nanosheet.}
	\end{figure}

	\subsection{Overall Architecture of iPREFER}
	
	The overall framework of the parameter extractor is depicted in Figure 2 and is primarily divided into four modules: Input Data, Feature Extractor, compact Model Parameter Extractor, and Model Verification.
	
	\begin{table}[h]
		\setstretch{1} 
		\caption{\textbf{\small{Key Parameters of 5nm Nanosheet}}}
		\label{tab:freq}
		\resizebox{0.6\linewidth}{!}{
			\begin{tabular}{cc}
				\toprule
				Key Parameters                           & Value        \\ 
				\midrule
				Lg                                       & $12nm$         \\
				High-k metal gate thickness              & $0.9nm$        \\
				Nanosheet thickness                      & $5nm$          \\
				The light doping in channel              & $10^{15}cm^{-3}$         \\
				The maximum doping in the source/drain   &  $10^{20}cm^{-3}$            \\
				The same vertical sheet to sheet spacing & $10nm$             \\
				The top/middle/bottom nanosheet widths   & $17/18.5/20nm$ \\ 
				\bottomrule
		\end{tabular}}
		\vspace{-10pt} 
	\end{table}
	
	The Input Data module encompasses two main components. Firstly, data for training the neural network involves random modifications to the compact model parameters, followed by Monte Carlo simulations using the SPICE simulator, resulting in 25,000 sets of Ids-Vgs (Low Vds), Cgg-Vgs (High Vds), and Cgg-Vgs (Low Vds) simulation curves. Secondly, TCAD simulation data is employed for validating the effectiveness of the parameter extractor. This dataset is based on TCAD simulation data from a 5nm nanosheet device, incorporating a ±10\% variation in LG and EOT to validate the high-precision parameter extraction for device variability.
	
	The next module is the Curve Feature Extraction module, responsible for extracting relevant features from the input curves. It extracts features corresponding to the IV and CV simulation curves of the input device, conveying the obtained curve features to the subsequent Model Parameter Extractor.
	
	The third module is the Model Parameter Extractor, which employs a core Artificial Neural Network (ANN). Through extensive training data, the ANN learns the mapping relationship between the device's IV and CV curve features wherever from the TCAD or experimental measurements and the compact model of the device.
	\begin{figure*}[]
		\centering
		\includegraphics[width=0.95\linewidth]{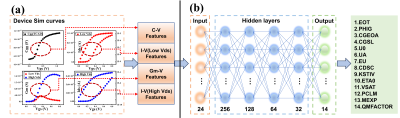}
		\caption{\textbf{\small{Diagram of iPREFER's Structure and Details: (a) Specific Structure of the Feature Extractor, (b) Structure Details of the Parameter Extraction Neural Network.}}}
		\vspace{-15pt}
		\Description{the 3D structural diagram and cross-sectional view of a 5nm nanosheet.}
	\end{figure*}
	
	Lastly, the Model Verification module evaluates the parameter extraction accuracy by inputting TCAD data into the pre-trained parameter extractor. This process involves comparing the SPICE simulation curves of the compact model with the initial TCAD simulation data, calculating the accuracy of the model parameter extraction.
	
	These four modules constitute the main structure of our proposed intelligent parameter extractor. Through this framework, a rapid and high-precision conversion from TCAD models to SPICE compact models can be achieved.
	\vspace{-8pt}
	\section{iPREFER: Details and Structure}
	
	This chapter provides a detailed overview of iPREFER, covering the details of the feature extractor, feature selection, and the specific structure of the deep learning network for parameter extraction.
	
		\begin{table}[h]
			\caption{\textbf{\small{Main Features Extracted by iPREFER from Different Curves}}}
		\resizebox{\linewidth}{!}{
			\begin{tabular}{|c|c|c|c|c|c|}
				\hline
				\textbf{Sim Curve}                                                                                 & \textbf{Feature}    & \textbf{Describe}                &\textbf{Sim Curve}                                         & \textbf{Feature}    & \textbf{Describe}         \\ \hline
				\multirow{6}{*}{\begin{tabular}[c]{@{}c@{}}Cgg-Vgs\\ Low Vds\\ Vgs(0V-1.1V)\end{tabular}} & Cgg\_max   & Max Cgg                 & \multirow{6}{*}{\begin{tabular}[c]{@{}c@{}}Ids-Vgs\\ High Vds\\ Vgs(0V-0.8V)\end{tabular}} & Vth2       & Threshold Voltage   \\ \cline{2-3} \cline{5-6} 
				& Cgg\_min   & Min Cgg                 &                                                                                            & SS2        & Subthreshold Swing  \\ \cline{2-3} \cline{5-6} 
				& V\_mid     & Voltage at Cgg Midpoint &                                                                                            & Ion2       & On-Current          \\ \cline{2-3} \cline{5-6} 
				& Cgg\_inte  & Integral of Cgg-Vgs     &                                                                                            & Ioff2      & Off-Current         \\ \cline{2-3} \cline{5-6} 
				& Cgg\_rms   & RMS of Cgg-Vgs          &                                                                                            & Ids\_inte2 & Integral of Ids-Vgs \\ \cline{2-3} \cline{5-6} 
				& Cgg\_slope & Cgg-vgs slope at Vgs=0  &                                                                                            & Ids\_rms2  & RMS of Ids-Vgs      \\ \hline
				\multirow{6}{*}{\begin{tabular}[c]{@{}c@{}}Ids-Vgs\\ Low Vds\\ Vgs(0V-0.8V)\end{tabular}} & Vth1       & Threshold Voltage       & \multirow{3}{*}{\begin{tabular}[c]{@{}c@{}}Gm-Vds\\ Low Vds\\ Vgs(0V-0.8V)\end{tabular}}   & Gm\_max1   & Max Gm              \\ \cline{2-3} \cline{5-6} 
				& SS1        & Subthreshold Swing      &                                                                                            & Gm\_inte1  & Integral of Gm-Vgs  \\ \cline{2-3} \cline{5-6} 
				& Ion1       & On-Current              &                                                                                            & Gm\_rms1   & RMS of Gm-Vgs       \\ \cline{2-6} 
				& Ioff1      & Off-Current             & \multirow{3}{*}{\begin{tabular}[c]{@{}c@{}}Gm-Vds\\ High Vds\\ Vgs(0V-0.8V)\end{tabular}}  & Gm\_max2   & Max Gm              \\ \cline{2-3} \cline{5-6} 
				& Ids\_inte1 & Integral of Ids-Vgs     &                                                                                            & Gm\_inte2  & Integral of Gm-Vgs  \\ \cline{2-3} \cline{5-6} 
				& Ids\_rms1  & RMS of Ids-Vgs          &                                                                                            & Gm\_rms2   & RMS of Gm-Vgs       \\ \hline
		\end{tabular}}
	\end{table}
	
	\subsection{Feature Extraction}
	The main structure of the feature extractor is depicted in Figure 4(a). Feature extraction is targeted at four sets of electrical characteristic curves of the device, aiming to extract relevant curve features. The extracted features include physical characteristics of the device, such as threshold voltage, sub-threshold swing, on-current, and off-current. Additionally, statistical and shape-related features of the curves, such as the integral, derivative, maximum, minimum, and square root of curve data, are also extracted. Table 2 enumerates the 24 key features extracted by iPREFER, with each feature corresponding to different parameters in the compact model. For instance, Vth corresponds to parameters like ETA0 and PHIG, while SS corresponds to parameters like CIT and CDSC. Through curve feature extraction, the abundance of data and less distinct features in the curve information is transformed into sensitive and rapidly quantifiable information regarding changes in compact model parameters.
	
	The core of iPREFER lies in extracting crucial physics and data mining from the electrical characteristic curves using the feature extractor. This allows the prediction of compact model parameters through a simpler neural network, yielding higher accuracy. The feature extractor transforms the originally complex input data into a set of concise and strongly correlated features with model parameters. This significantly accelerates iPREFER's convergence speed and enhances prediction accuracy, facilitated by a feature extractor grounded in both physical and statistical information.
	
	\subsection{ Parameter Extraction Network}
	Figure 4(b) illustrates the specific structure and details of the parameter extraction neural network, comprising a simple Artificial Neural Network (ANN) with four hidden layers. To address the vanishing gradient issue in deep networks, the SWISH activation function is employed\cite{swish}. The deep learning network takes as input normalized electrical characteristics of device curves and outputs 14 compact model parameters corresponding to the device's electrical characteristics. This essentially covers the main model parameters required in the manual parameter extraction process. It is crucial to emphasize that the model's output parameters can be adjusted based on specific requirements by adding or removing curve features strongly correlated with the model parameters to the training data.
	
	The underlying nature of the BSIM-CMG compact model involves a complex current-voltage equation model, where the compact model parameters represent a series of parameters within these equations. Curve features are obtained through curve calculations, establishing a clear correspondence between curve features and the compact model. The network's purpose is to characterize the mapping relationship between curve features and compact model parameters.
	
	In contrast to traditional deep learning methods that typically train using IV (current-voltage) and CV (capacitance-voltage) curves as direct input data, iPREFER streamlines the process by focusing on the physical features of curves. This approach significantly simplifies the complexity of traditional deep learning methods, eliminating the need to separately train CV and IV model parameters, ensuring consistency and accuracy in parameter extraction.
	
	The training process of iPREFER involves leveraging SPICE simulation tools to perform Monte Carlo simulations by modifying the corresponding compact model parameters, resulting in a dataset of 25,000 IV and CV curves. The device curves, after feature extraction, selection, and normalization, serve as the training data for the neural network. After successfully training the neural network, the effectiveness of the parameter extractor is validated using simulation curves from 5nm nanosheet TCAD devices. It is essential to note that during the generation of training data, the compact model parameters corresponding to TCAD device simulation curves should fall within randomly generated parameter ranges.

		\vspace{-10pt}

	\section{Testing Results by TCAD Data}
	
	This chapter will begin by presenting the results of inputting benchmark TCAD data into the parameter extractor to obtain the corresponding compact model. Subsequently, the chapter will delve into the testing results of the parameter extractor when TCAD data's EOT and LG are varied by ±10\%.
	
		\vspace{-10pt}
	\subsection{Testing Results of Baseline TCAD Data }
	
	After completing the training, we utilized the baseline TCAD data as input for the iPREFER tool to test its capability in accurately extracting corresponding compact model parameters. Following the extraction of the compact model, we employed the SPICE simulation tool to obtain its simulation curves, and the obtained compact model simulation curves were compared with the corresponding TCAD simulation data, as shown in Figure 5. 
	
	\begin{figure}[h]
		\centering
		\includegraphics[width=0.95\linewidth]{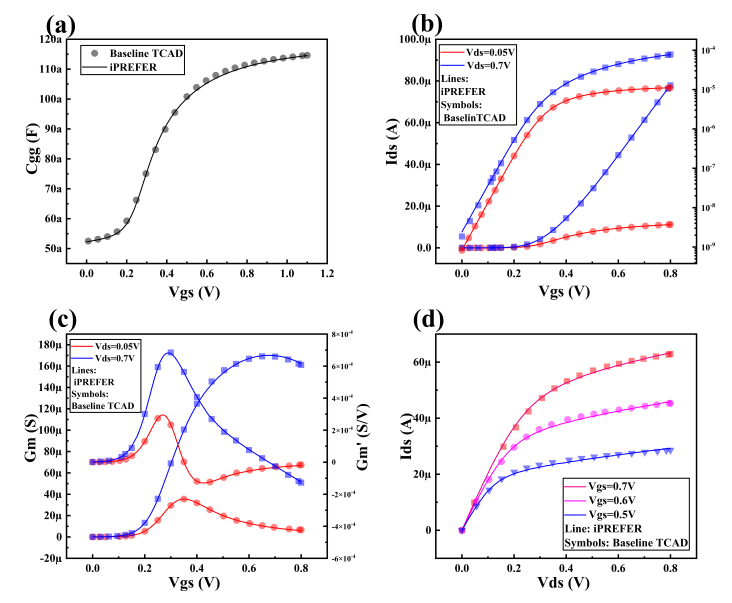}
		\caption{\textbf{\small{Model Testing Results Extracted by iPREFER from Baseline TCAD Data: (a) Cgg-Vgs, (b) Ids-Vgs at Vds=0.05V and 0.7V, (c) Gm-Vgs and Gm'-Vgs at Vds of 0.05V and 0.7V, and (d) Ids-Vgs at Vgs of 0.5V, 0.6V, and 0.7V.}}}
		\Description{the 3D structural diagram and cross-sectional view of a 5nm nanosheet.}
	\end{figure}

	\begin{figure}[h]
		\centering
		\includegraphics[width=0.95\linewidth]{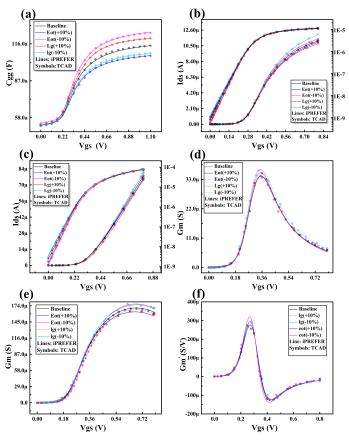}
		\caption{\textbf{\small{Model Testing Results Extracted by iPREFER from Variable TCAD Data: (a) Cgg-Vgs, (b) Ids-Vgs at Vds=0.05V, (c) Ids-Vgs at Vds=0.7V, (d) Gm-Vgs at Vds=0.05V, (e) Gm-Vgs at Vds=0.7V, and (f) Gm'-Vgs at Vds=0.05V.}}}
		\Description{the 3D structural diagram and cross-sectional view of a 5nm nanosheet.}
	\end{figure}

	In Figure 5(a), the RMS error of the Cgg-V curve is 0.42\%. In Figure 5(b), the RMS error of the Ids-Vgs(Vds=0.05V) curve is 1.28\%, with a RMS error of 0.32\% in the sub-threshold voltage region. The RMS error of the Ids-Vgs(Vds=0.7V) curve is 2.6\%, with a fitting error of 0.95\% in the sub-threshold voltage region.To further validate the accuracy of the extracted compact model, Figures 5(c) depicts the fitting results for the first and second derivatives of Ids-Vgs, respectively. In Figure 5(d), the RMS error of the Ids-Vds curve is 1.58\%. The fitting accuracy of the main IV and CV curves essentially reaches the ultimate precision for fitting nanosheet devices with the BSIM-CMG compact model.  The iPREFER demonstrates sufficient accuracy in extracting the compact model from baseline TCAD data, as evidenced by the high precision fitting of higher-order derivatives. This fully demonstrates the capability of iPREFER to accurately extract the compact model from baseline TCAD data.

	\begin{figure}[h]
		\centering
		\includegraphics[width=0.75\linewidth]{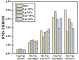}
		\caption{\textbf{\small{Accuracy Assessment of iPREFER with Variable TCAD Data.}}}
		\Description{the 3D structural diagram and cross-sectional view of a 5nm nanosheet.}
	\end{figure}
	
	\vspace{-10pt}
	
	\subsection{Testing Results of Variable TCAD Data }	
	Variability and stochastic fluctuations in advanced process devices are crucial factors affecting device electrical characteristics and circuit performance. Transforming process inconsistencies into variations in compact model parameters has been a challenge for traditional parameter extraction tools. To validate the high-precision parameter extraction capability of iPREFER for device process variability, we conducted an in-depth analysis on two key parameters, Lg and EOT, in different dimensions. By individually modifying EOT by ±10\% and LG by ±10\%, the corresponding data were input into iPREFER to obtain the corresponding compact models. Through SPICE simulation, compact model simulation curves were generated and compared with the initial TCAD data in fitting graphs shown in Figure 6. Figure 7 illustrates the RMS errors for these five sets of curves. The average RMS errors are 0.47\% for Cgg-Vgs, 1.44\% for Ids-Vgs(Vds=0.05V), 2.47\% for Ids-Vgs(Vds=0.7V), 3.98\% for Gm-Vgs(Vds=0.05V), and 3.59\% for Gm-Vgs(Vds=0.7V). These results demonstrate that iPREFER accurately extracts compact models for variable TCAD data, and the parameter extraction accuracy is comparable to baseline TCAD data.

	\section{Advancements of iPREFER}

	In this chapter, we will elaborate on the advancements of iPREFER, sequentially discussing its high performance, enhanced parameter extraction accuracy and generalization capability, and its attributes as a flexible and universally applicable parameter extraction method.
	
	\subsection{ Exceptional Parameter Extractor Performance }	
	Prior to training the network, iPREFER extracts features from device IV and CV curves, effectively reducing the overall model size of the parameter extraction network. This enables iPREFER to precisely extract compact models for nanosheet devices using a smaller ANN network. Table 3 presents a comparative analysis of iPREFER against other mainstream algorithms. iPREFER demonstrates superior performance metrics, with fewer networks, fewer neurons, shorter training times, smaller training datasets, quicker dataset generation times, and higher parameter extraction accuracy compared to other AI-based parameter extraction algorithms. iPREFER outperforms other AI-based parameter extraction algorithms across major performance indicators.
	
	\begin{table}[h]
		\caption{\textbf{\small{Comparison of Parameters with Other Algorithms}}}
		\centering
		\resizebox{0.85\linewidth}{!}{
			\begin{tabular}{lccc}
				\toprule
				\textbf{Parameter} & \textbf{iPREFER} & \textbf{DLPE\cite{dlpe}} & \textbf{DLFPE\cite{dlfpe}} \\
				\midrule
				Test Technology Platform & 5nm nanosheet & 10nm FinFET & 14nm FinFET \\
				Extracted Parameters & 14 & 12 & 14 \\
				Baseline TCAD Ids-Vgs Average RMS Error & 2.0\% & 6.1\% & 3.4\% \\
				Variable TCAD Ids-Vgs Average RMS Error & 2.2\% & 14.2\% & Not supported \\
				Dataset Generation Time & 654s & $\sim$900s & 1800s \\
				Training Time & 345.94s & $\sim$600s & $\sim$3300s \\
				Dataset Size & 25000 & 50000 & 100000 \\
				Number of Networks & 1 & 2 & 17 \\
				Number of Neurons & 340 & 2400 & $\sim$2500 \\
				Support for Variable Process Parameters & Yes & Yes & No \\
				Support for Extended Parameters & Yes & No & Yes \\
				\bottomrule
		\end{tabular}}
		\label{tab:parameter_comparison}
	\end{table}
	
	\vspace{-10pt}
	
	\begin{figure}[h]
		\centering
		\includegraphics[width=0.95\linewidth]{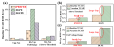}
		\caption{\textbf{\small{Comparison of Testing Errors between iPREFER and DLPE/DLFPE: (a) Precision Comparison on Baseline TCAD Data, (b) Testing Error Disparity beyond the Threshold Voltage Range, (c) Testing Error Discrepancy across the Entire Vgs Range.}}}
		\Description{the 3D structural diagram and cross-sectional view of a 5nm nanosheet.}
	\end{figure}
	\vspace{-10pt}
	\subsection{Higher Accuracy and Generalization Capability }
	iPREFER is designed for the 5nm nanosheet device platform, while DLPE pertains to 10nm FinFET and DLFPE to 14nm FinFET. iPREFER exhibits advanced parameter extraction capabilities for more sophisticated devices than other algorithms. Figure 8(a) illustrates the parameter extraction accuracy comparison between iPREFER and other algorithms when using Baseline TCAD data as input. Due to the complexity of the nanosheet's capacitance model, iPREFER's capacitance parameter extraction accuracy is lower than DLPE\cite{dlpe} and DLFPE\cite{dlfpe}, but it excels in the parameter extraction all extremely precise of major Ids-Vgs and Gm-Vgs curves. Figures 9(b) and 9(c) respectively showcase the accuracy comparison between iPREFER and DLPE under Baseline TCAD data and variable TCAD data testing, spanning the full Vgs range and the range greater than the threshold voltage. Compared to DLPE, iPREFER demonstrates consistent parameter extraction accuracy for variable TCAD and baseline TCAD, indicating stronger learning capabilities and superior generalization capabilities compared to other deep learning-based parameter extraction algorithms.
	\subsection{Versatile Frameworks for Parameter Extraction}
	iPREFER serves as a universal parameter extraction method, imposing no strict requirements on input data. By extracting corresponding physical and data features from the device's electrical characteristic curves and employing these features to predict compact models, iPREFER is not limited to BSIM-CMG compact models. Other compact models for various devices can also leverage this parameter extraction method to achieve high-precision and efficient parameter extraction. However, the accuracy of parameter extraction is intricately linked to the quality and quantity of selected curve features, relying on the physical equations of compact models for feature selection and extraction.
	
	\section{Conclusion}
	To facilitate the critical transition from TCAD device models to device compact models within the DTCO workflow, this paper introduces for the first time an intelligent parameter extraction tool called iPREFER, based on both physical and data features. iPREFER employs feature extraction from the electrical characteristics curves of devices and constructs a high-speed, high-precision, and robust parameter extraction framework using a straightforward neural network structure. Validation of iPREFER's parameter extraction accuracy was conducted using the electrical simulation curves of 5nm nanosheet TCAD devices. The obtained accuracy surpasses that of existing deep learning algorithms, essentially reaching the precision limits achieved through manual tuning. iPREFER is equally applicable to the extraction of compact model parameters for other advanced process devices, extending beyond the scope of 5nm nanosheets. This pioneering integration of an AI-accelerated compact model parameter extraction tool into the DCTO process significantly accelerates the transfer of device process parameters to circuit performance. This acceleration, in turn, expedites the iterative process of refining device process parameters for enhancing circuit design. The incorporation of iPREFER into the entire DTCO workflow holds substantial implications.


\bibliographystyle{IEEEtran}
\bibliography{IEEEabrv,sample-base.bib}


\end{document}